\documentclass[12pt]{JHEP}

\usepackage{amsmath2000,amssymb,euscript,array,cite} 

\newcommand{\startappendix}{
\setcounter{section}{0}
\renewcommand{\thesection}{\Alph{section}}}
\newcommand{\Appendix}[1]{
\refstepcounter{section}
\begin{flushleft}
{\large\bf Appendix \thesection: #1}
\end{flushleft}}

\usepackage{epsfig}

\def\N{{\cal N}}

\def\Tr{{\rm Tr}\,}
\def\sst{\scriptscriptstyle}

\def\SU{\text{SU}}
\def\U{\text{U}}

\def\SL{\text{SL}}

\def\Dbarslash{\,\,{\raise.15ex\hbox{/}\mkern-12mu {\bar\D}}}
\def\Dslash{\,\,{\raise.15ex\hbox{/}\mkern-12mu \D}}
\def\delslash{\,\,{\raise.15ex\hbox{/}\mkern-9mu \partial}}
\def\delbarslash{\,\,{\raise.15ex\hbox{/}\mkern-9mu {\bar\partial}}}

\def\VEV#1{\left\langle #1\right\rangle}

\newcommand{\MAT}[1]{\begin{pmatrix} #1\end{pmatrix}}
\newcommand{\EQ}[1]{\begin{equation} #1 \end{equation}}
\newcommand{\AL}[1]{\begin{subequations}\begin{align} #1
\end{align}\end{subequations}}
\newcommand{\SP}[1]{\begin{equation}\begin{split} #1 \end{split}\end{equation}}

\def\np#1#2#3{{\it Nucl. Phys.} {\bf B#1}, #2 (#3)}

\def\prl#1#2#3{{\it Phys. Rev. Lett.} {\bf #1}, #2 (#3)}

\def\jhep#1#2#3{{\it JHEP} {\bf #1}, #2 (#3)}

\title{World-sheet Instantons via the Myers Effect and  
$\N=1^*$ Quiver Superpotentials} 
\author{Timothy J.~Hollowood
and S.~Prem~Kumar\\
Department of Physics, University of Wales Swansea,
Swansea, SA2 8PP, UK\\
E-mail: {\tt t.hollowood@swan.ac.uk},
{\tt s.p.kumar@swan.ac.uk}}
\preprint{ SWAT-}
\abstract{In this note we explore the stringy interpretation of
non-perturbative effects in $\N=1^*$ deformations of the $A_{k-1}$
quiver models. For certain types of deformations we argue that the
massive vacua are described by $Nk$ fractional D3-branes at the
orbifold polarizing into $k$ concentric 5-brane spheres each carrying
fractional brane charge. The polarization of the D3-branes {\it
induces} a polarization of D-instantons into string world-sheets
wrapped on the Myers spheres. We show that the superpotentials in
these models are indeed generated by these world-sheet instantons. We
point out that for certain 
parameter values the condensates yield the exact superpotential for a
relevant deformation of the Klebanov-Witten conifold theory.}

\begin{document}

\section{Introduction}

The AdS/CFT correspondence in its principal form proposes a duality
between large-$N$ (super)conformal theories (SCFT's) in four dimensions and
Type IIB superstring 
theory on $AdS_5\times X^5$.
This duality may be used as a 
starting point to obtain further dualities 
between large-$N$ confining gauge theories in four dimensions
and IIB superstring theory. 
In this context much attention has been focussed on relevant
perturbations of the $\N=4$ $\SU(N)$ theory, namely, the so-called $\N=1^*$
theory \cite{gppz,nickprem,polstr, oferandus, nick} which has vacua where the
theory
confines and generates a mass gap, exhibiting an extremely rich phase
structure.  

More recently in \cite{us} the vacuum and
phase structure of $\N=1^*$ perturbations of the $A_{k-1}$ quiver gauge
theories ($\N=2$ SCFT's with gauge group
$\SU(N)^k$ and bi-fundamental matter) have been explored in detail revealing
even richer infrared dynamics. One of the most remarkable aspects of this 
class of relevant perturbations of the $\N=4$ and $\N=2$ quiver theories is
that the holomorphic sector of the resulting $\N=1$ theories can be computed
exactly as a function of $N$ and the gauge couplings (or marginal
parameters). In particular, the exact superpotential for the mass-deformed
$\N=4$ theory was 
obtained in \cite{nick} while the superpotential for the $\N=1^*$ quiver
theories was derived in \cite{us}. The aim of this note is to elucidate, from a
stringy point of view, the non-perturbative effects that generate these
superpotentials and to point out the ingredients of the string duals of the 
$\N=1^*$ quiver theories. For a certain class of deformations 
we demonstrate that detailed features of the
superpotentials can be understood simply via the Myers effect
\cite{polstr, myers} operating on 
D-instantons turning them into
string world-sheet instantons. The superpotentials are generated by string
instantons wrapping flux-supported two-spheres obtained by
polarization of D3-branes into 5-branes. 

The exact superpotentials and condensates \cite{nick, us} for the
$\N=1^*$ theories 
are elliptic functions of the marginal parameters (the couplings
of the individual gauge groups in the quiver) $\tau_i =4\pi
i/g^2_i+\theta_i/2\pi$ with the IIB string coupling $\tau=i/g_s+C^0$
being the modular parameter. 
These have a simple semiclassical
$(\tau_i\rightarrow i\infty)$ expansion which in a generic vacuum can be
interpreted as contributions from objects carrying a certain topological
charge with respect to each gauge group factor of the
quiver. In particular for vacua in the Higgs phase these are simply instanton
contributions while for vacua in the confining phase the corresponding
expansions appear to encode contributions from configurations
carrying fractional topological charge within
each gauge group factor of the quiver.

In this note we will show how the non-perturbative physics 
controlling the holomorphic sector of the
massive vacua of ${\cal N}=1^*$ quiver theories can be understood in
a relatively simple way in the IIB theory via D3-branes at
an orbifold in the presence of non-trivial fluxes corresponding to the 
${\cal N}=1^*$ deformations. By analysing appropriate D$(-1)$-D3-brane
configurations at weak string coupling and
finite $N$, we outline ``selection rules'' that explain the detailed 
features of instanton
contributions that generate the exact superpotentials in the
Higgs vacua of ${\cal N}=1^*$ theories. By $S$-duality we obtain the
corresponding pictures for the confining vacua of these theories as
well. 

We focus mainly on $\N=1^*$ perturbations that respect the 
symmetry under exchanges of the $\SU(N)$ factors in the quiver. In
particular this implies equal $\N=2$ SUSY preserving masses $m$ for
the $k$ bi-fundamental hypermultiplets and equal $\N=1$ SUSY preserving
masses $\mu$ for the $k$
adjoint chiral multiplets via the following tree-level superpotential
perturbation:
\EQ{\Delta W =-{1\over g^2_{YM}}\sum_{i=1}^k 
\left(m\Tr[\Phi_{i,i+1}\Phi_{i+1,i}]+\mu\Tr\Phi_i^2
\right).\label{tree}}
The exact low-energy effective superpotential is then,
\EQ{W_{\rm eff}=-{\mu\over g^2_{YM}}\sum_{i=1}^k\VEV{\Tr\Phi_i^2}.}

In the Higgs vacua of the resulting $A_{k-1}$ $\N=1^*$ theory we find that the
(fractional) D3-branes at an $A_{k-1}$ orbifold polarize into
precisely $k$
D5-branes via the dielectric effect \cite{polstr, myers}  by acquiring
non-commuting positions. We also argue that the Higgs vacuum
configurations suggest that each of these $k$ polarized D5-branes must
be homologous to a topologically distinct $S^2$ cycle of the resolved
orbifold. An interesting aspect of the vacuum solutions is that they
spontaneously break the symmetry of the theory under exchanges of the
$\SU(N)$ factors. From the point of view of the IIB string picture this
implies that twisted sector modes are turned on. 

We find, subject to certain caveats, that
the vacuum equations of the D-instanton gauge theory in the
presence of these expanded or polarized D5-branes have solutions
that allow only specific numbers of
D-instantons to contribute to the expanded brane configuration in a given
vacuum. This explains, for instance in the $k=1$ case, why instantons
contribute only in multiples of $N$ in the Higgs vacuum of the $\SU(N)$, mass
deformed ${\cal N}=4$ theory. More elaborate selection rules emerge
for the $A_{k-1}$ quiver theories in general. Importantly, the
polarization of the fractional D3-branes as a response to the
background flux, {\it induces} a polarization of the ``dissolved''
D-instantons into wrapped D-string world-sheets whose action is exactly
computable at weak string coupling. Thus even at arbitrarily weak
coupling and finite $N$, non-perturbative effects in the Higgs vacua
can be understood in the IIB D-brane setup as world-sheet instanton
contributions. This provides a ``microscopic''
picture of the world-sheet instantons of \cite{nickprem,polstr}
which are responsible for non-perturbative effects in the IIB string
duals in the large-$N$ limit of $\N=1^*$ theory at large 't Hooft
coupling. $S$-duality turns the Higgs phase into a confining phase while
simultaneously turning the wrapped D-string into a wrapped F-string. In
this sense it is natural to think of non-perturbative effects in the
confining vacua as being due to F-string world-sheets wrapping
NS5-branes obtained by polarizing the fractional D3-branes at the
orbifold.\footnote{Strictly speaking the expansion of
the superpotentials in terms of world-sheet instantons in the
confining vacuum is valid in the regime of large `t Hooft coupling
$g_sN\gg 1$ which coincides with the SUGRA regime.}

A nontrivial ingredient in the above interpretation is that 
world-sheet instanton actions wrapping the polarized 5-branes at the
orbifolds are expected to be sensitive to twisted sector fields. This
is because the polarized 5-brane-spheres are homologous to the
two-cycles at the resolved orbifold. Hence wrapped F-string world-sheet actions
should be accompanied by phase factors proportional to the integral of
the 
$NS$ or $RR$ two-form over the associated two cycles.
In fact precisely such phases appear in the confining vacuum in a
strong-coupling expansion
of the  
superpotential via terms  proportional to powers of $\exp (i\int
B_{NS}/2\pi\alpha^\prime)$. Thus, the field theory in 
the confining vacua appears to naturally encode stringy effects 
in a strong coupling expansion.

For the sake of simplicity, for the most part we restrict our
attention to explicit results in the $A_1$ or $\SU(N)\times\SU(N)$
quiver model. Our results can be extended straightforwardly to the
$A_{k-1}$ models for general $k$. 

We also comment on the exact superpotential for the $\N=1^*$
deformation of the $\SU(N)\times\SU(N)$ theory where the two adjoint
chiral multiplets have masses $\mu_1=-\mu_2=\mu$. This theory may be
viewed as a relevant deformation of the Klebanov-Witten conifold
theory \cite{Klebanov:1998hh} 
\EQ{\Delta W=-{1\over g^2_{YM}} \left(
\mu\Tr\Phi_1^2-\mu\Tr\Phi_2^2+
m\Tr\Phi_{1,2}\Phi_{2,1}
+m\Tr\tilde{\Phi}_{2,1}\tilde\Phi_{1,2}\right)\ .}
For large hypermultiplet masses $m\gg \mu$, this is softly broken $\N=2$
SUSY Yang-Mills with massive  bi-fundamental hypermultiplets and gauge group
$\SU(N)\times\SU(N)$. For $m\ll \mu$, however, it can be thought of as a
deformation of the Klebanov-Witten theory which in turn may be viewed
as a perturbation of the $A_1$ $\N=2$ quiver theory. Based on the analysis of
\cite{us}, remarkably, the massive phases and massive vacuum structure
and condensates of the
resulting theory are identical to that of the theory with tree-level
superpotential (\ref{tree}). This allows us to evaluate the
superpotential in the massive vacua exactly.

\section{${\cal N}=1^*$ quiver models:  a brief review} 

The $A_{k-1}$ quiver gauge theories with $\N=2$ supersymmetry 
describe the low-energy dynamics on the 
world-volume of $N$ D3-branes at a ${\mathbb Z}_k$ orbifold
singularity in an $A_{k-1}$ ALE space \cite{dougmoore}. 
This theory is conformal and has $k$ exactly marginal couplings, one for each
gauge group factor
\EQ{\tau_i={4\pi i\over g^2_i}+{\theta_i\over 2\pi};{\quad}i=1,2,\ldots k.}
We may also naturally define an `overall' gauge coupling 
\EQ{\tau=\sum_{i=1}^k\tau_i={4\pi i\over g^2_{YM}}+{\theta\over
2\pi}\label{overall}}
to be
thought of as the gauge coupling for the diagonal $\SU(N)$ factor and 
is also the coupling constant of the IIB string $\tau=i/g_s+C^{0}$.
In addition to $k\;$ $\N=2$ vector multiplets $(W_\alpha^i,\Phi_i)$,
the quiver theory has $k$ 
bi-fundamental hypermultiplets $\{\Phi_{i,i+1},\Phi_{i+1,i}\}$
transforming in the $(({\bf N},\bar{\bf
N}), (\bar{\bf N},{\bf N}))$ representation of the $i^{\rm th}$ and $i+1^{\rm
th}$ $\SU(N)$ 
factors in the quiver ($i=1\ldots k$). 
The conformal ${\cal N}=2$ quiver theory also has a tree-level superpotential
\EQ{W={2\sqrt 2\over g^2_{YM}}\sum_{i=1}^k 
\Tr\Phi_i\big(\Phi_{i,i+1}\Phi_{i+1,i}-\Phi_{i,i-1}\Phi_{i-1,i}\big).\label{suptree}}

The ${\cal N}=1^*$ quiver theory is obtained by adding to the
superpotential (\ref{suptree}) generic ${\cal
N}=1$ preserving masses $\mu_i$ for the $k$ adjoint chiral
multiplets, and generic masses $m_i$ for the bi-fundamental
hypermultiplets, 
\EQ{\Delta W =-{1\over g^2_{YM}}\sum_{i=1}^k 
\left(m_i\Tr\big(\Phi_{i,i+1}\Phi_{i+1,i}\big)+\mu_i\Tr\Phi_i^2
\right).\label{pert}}
For generic mass parameters, the determination of the vacuum structure
is somewhat involved \cite{us}, but the final picture
that emerges is rather simple and nicely generalizes the story for the
mass-deformed ${\cal N}=4$ theory. In particular the theory has
discrete vacua with mass gap, and 
\EQ{{\rm No.}\; {\rm of}\; {\rm massive}\; {\rm vacua}\;=N^{k-1}\sum p,\;\;{\rm
where}\;p\;{\rm divides}\;N.}
Vacua labelled by a divisor $p$ arise from classical configurations
that preserve an $\SU(p)^k$ gauge symmetry with ${\cal N}=1$ SUSY. 
The theory has a ${\mathbb Z}_N$ symmetry that resides in the center
of diagonal $\SU(N)$ gauge transformations. Thus as in the case of the
mass-deformed ${\cal N}=4$ theory, the massive phases of the theory in
these vacua are classified 
by the order $N$ subgroups of ${\mathbb Z}_N\times{\mathbb Z}_N$
\cite{polstr,thooft,donwitt}. 
Each massive phase is realized with a multiplicity
$N^{k-1}$.
Specifically, there are $N^k$ vacua where the theory is in a
(electric) confining phase (with $p=N$) and $N^{k-1}$ vacua where it is in
the Higgs phase ($p=1$). As in the case of the mass-deformed ${\cal
N}=4$ theory, $\SL(2,{\mathbb Z})$ duality on $\tau$ of the ${\cal N}=1^*$
quiver theory relates the theories in
the vacua with different phases, while simultaneously rescaling the individual
gauge couplings $\tau_i$. Actually, the conformal ${\cal N}=2$ quiver
model has an extended duality group \cite{wittm, Katz:1997eq,Lawrence:1998ja}
which acts by
permutation on all the massive vacua of the ${\cal N}=1^*$ quiver theory
\cite{us}. 

In this note we will focus attention primarily on the
Higgs and confining vacua of the ${\cal N}=1^*$ quiver models. Our
observations can be carried over with some modifications to include
all other vacua and phases of these models.

\section{The Higgs vacua in field theory \label{one}}

For the sake of simplicity we will first assume equal
bi-fundamental masses $m_i=m$ and equal masses for the adjoints
$\mu_i=\mu$. Many qualitative features of the physics
that we discuss, will not be affected by this simplifying choice.
Furthermore, we will use the vacua in the Higgs phase as our starting
point since these can be reliably understood at weak coupling $g^2_i
N/4\pi \ll 1$. The
action of dualities on the resulting picture will describe vacua in
other (confining) phases.

The classical superpotential of the ${\cal N}=1^*$ quiver theory can
then be recast as \cite{us}
\EQ{W={1\over g^2_{YM}}
\Tr\left(\Phi[\Phi^+, \Phi^-]-m\Phi^+\Phi^- -\mu\Phi^2\right)}
where $\Phi^\pm$ and $\Phi$ are $kN\times kN$ matrices into which the
$k$ bi-fundamentals and adjoints, respectively, can be conveniently amalgamated.
We choose an ordering for the elements so that the rows and columns
numbered $i,i+k,i+2k,\ldots i+(N-1)k$ are associated with the $i^{\rm th}$
$\SU(N)$ factor in the quiver. With this ordering in place the only
non-zero elements of $\Phi$ are in the positions $(u,u+nk)$ with $n$
an integer. Similarly the only non-zero elements of $\Phi^\pm$ are in
the positions $(u,u\pm1+nk)$ with $n$ an integer.

With our simplifying choice of mass parameters, the $F$-term
equations are simply the $\SU(2)$-commutation relations after a trivial
rescaling of the field variables:
\EQ{[\Phi,\Phi^\pm]=\pm m \Phi^\pm,\;\;\;\;\;\;
[\Phi^+,\Phi^-]=2\mu \Phi.\label{fterm}}
As usual solving $D$ and $F$ flatness conditions and modding out by
gauge transformations is equivalent to
solving the $F$-term equations modulo complexified gauge
transformations. The complex gauge transformations can be used up to
bring the matrices $\Phi, \Phi^\pm$ to a form where 
$\Phi\propto J_3$ and $\Phi^\pm$ have only off-diagonal non-zero elements in the
positions $(u,u\pm1)$ as is to be expected for step-up and step-down
generators $J^\pm$ of the $\SU(2)$ algebra. Massive vacua are associated with
special {\it reducible} representations of the $\SU(2)$ algebra.

It was shown in \cite{us} that for solutions to Eq. (\ref{fterm}) which
break the gauge group completely, {\it i.e.\/}~in the Higgs vacua, the matrices
$\Phi,\Phi^\pm$ form
reducible representations of the $\SU(2)$ algebra made up of precisely 
$k$ irreducible blocks: 
\EQ{
(\Phi,\Phi^\pm)=\MAT{   (\Phi_1,\Phi^\pm_1)_{\sst[\ell_1]\times[\ell_1]}
&&&\\ &(\Phi_2,\Phi^\pm_2)_{\sst[\ell_2]\times[\ell_2]}&&\\%
&&\ddots&\\
&&&(\Phi_k,\Phi^\pm_k)_{\sst[\ell_k]\times[\ell_k]}}\ ,
\label{Higgsol}
}
where $\sum\ell_r=Nk$ and each irreducible block forms a representation of the
$\SU(2)$
algebra satisfying the usual Casimir relation:
\EQ{
{\Phi_r^2\over m^2}+{1\over {2\mu m}}
(\Phi^+_r\Phi^-_r+\Phi^-_r\Phi^+_r)={(\ell_r^2-1)\over 4}.
\label{casimir}
}

However, not all such reducible representations break the gauge group
completely. When $k=1$ of course, there is only one
such solution which is the $N\times N$ irreducible representation of
$\SU(2)$ corresponding to the single Higgs vacuum of the mass-deformed
${\cal N}=4$ theory. For $k>1$, the situation is explained in
\cite{us}. We simply quote the results below.
\subsection{$\SU(N)\times\SU(N)$ theory}

In the simplest non-trivial case with $k=2$ {\it i.e.\/}~the $\SU(N)^2$
${\cal N}=1^*$ theory, Higgs vacuum solutions correspond to reducible
representations composed of 2 blocks with dimensions $\ell_1$ and $\ell_2$ with
$\ell_1+\ell_2=2N$ and $\ell_1=1,3,5,\ldots 2N-1$, thus yielding $N$
Higgs vacua. It is important to note that based on the ordering of
elements chosen above, the odd-numbered rows and columns are
associated to the first $\SU(N)$ factor, while the even-numbered
ones are associated to the second $\SU(N)$ factor. 
For general $k$ there are $N^{k-1}$ distinct classical solutions
that Higgs the gauge group completely.

\subsection{$\SU(N)^k$ theory}

For general $k$ the Higgs vacua may be classified as follows \cite{us}. They
correspond to partitions of the {\it ordered} set of $kN$
diagonal elements of $\Phi\propto J_3$ into precisely $k$ sets each of dimension
$\ell_k$
\footnote{Partitions into a larger number of sets gives rise to massless vacua
or vacua with
a classically unbroken non-Abelian gauge group.}
\SP{
&
\Big\{1,2,\ldots,k,1,2,\ldots,k,\ldots\ldots,1,2\ldots,k\Big\}\\
&\qquad\qquad\qquad\qquad\qquad
\longrightarrow\Big\{\underbrace{1,\ldots,i_1}_{{\EuScript A}_1}\Bigg|
\underbrace{i_1+1,\ldots,i_2}_{{\EuScript A}_2}\Bigg|
\ldots\ldots\Bigg|\underbrace{i_{n-1}+1,\ldots,
k}_{{\EuScript A}_k}\Big\}\ .
\label{part}
}
with $i_k=k$ and importantly
\EQ{
\Big\{i_1,i_2,\ldots,i_{k-1}\Big\}=\Big\{1,2,\ldots,k-1\Big\}\ .
\label{alpt}
}
It is important to note that two partitions $\{{\EuScript A}_r\}$ and
$\{{\EuScript A}^\prime_r\}$  are equivalent if the second partition
yields the first upon a simple re-labelling of the subsets ${\EuScript
A}^\prime_r$. They are related by the action of the Weyl group of $\SU(N)^k$.

The number of such distinct partitions can be 
determined easily. There is a single $i_r$ in
\eqref{part} associated to each of the $k$ gauge group factors. A
given $i_r$ can therefore be situated in one of $N$ places. However,
$i_k$ is fixed, so the total degeneracy of Higgs vacua is
$N^{k-1}$. 

\section{Type IIB picture }

The conformal ${\cal N}=2$ quiver theory arises as the low-energy
world-volume dynamics of $N$ D3-branes at a ${\mathbb Z}_k$ orbifold
singularity in an $A_{k-1}$ ALE space \cite{dougmoore}. We choose the
D3-brane world-volume  
to span the coordinates $x^0,\ldots,x^3$. The six-dimensional
space transverse to the the D3-branes is ${\mathbb R}^2\times({\mathbb
R}^4/{\mathbb Z}_k)$. We choose the $(x^4,x^5)$ coordinates to
parameterize the transverse plane left fixed by the orbifolding:
\EQ{x^4,x^5\rightarrow x^4,x^5;\quad\quad
x^6+ix^7\rightarrow e^{2\pi i\over k}(x^6+ix^7);\quad\quad
x^8+ix^9\rightarrow e^{-{2\pi i\over k}}(x^8+ix^9).}

When the D3-branes sit at the orbifold
singularity, {\it i.e.\/}~on the fixed plane, they are free to
fractionate. In general they split into $Nk$ fractional D3-branes
which are secretly 5-branes wrapped on one of the $k-1$ collapsed $S^2$'s at
the orbifold singularity \cite{Polchinski:1996ry, Diaconescu:1997br}.
The $(x^4,x^5)$ separations between the fractional D3's appear as the
$k$ adjoint scalars parameterizing the $(N-1)k$-dimensional Coulomb
branch of the quiver model with $\SU(N)^k$ gauge group. Strictly
speaking the gauge 
group is $\U(N)^k$, but the overall $\U(1)$ simply decouples from the
dynamics while all the remaining $\U(1)$'s are IR-free and freeze 
out at low-energies.

The IR-free $\U(1)$ factors correspond to the relative positions
of the centers of masses of the $k$ groups of fractional D3-branes on the
fixed $(x^4,x^5)$ plane. Non-zero hypermultiplet masses $m_i$
satisfying $\sum m_i=0$ may be introduced by displacing the
centers of masses of these $k$ groups. The ${\cal N}=1^*$ theories
have masses $\mu_i$ for the $k$ adjoint-valued scalars. As long as the
$\mu_i$ satisfy $\sum \mu_i=0$ they may be directly
related to  blow-up modes of the orbifold singularity
\cite{Gubser:1998ia}. In particular switching on such masses produces resolved 
spaces which are fibrations of a smooth ALE on the $(x^4,x^5)$ plane.

In the IIB picture, the overall adjoint mass $\sum \mu_i =\mu$ appears
when certain components of the 3-form fluxes $F^{(3)}={\;}^*dC^{(6)}_{RR}$ and 
$H^{(3)}=dB^{(2)}_{NS}$ are turned
on. The condition $\sum m_i = 0$ can also 
be relaxed by switching on certain components of the 3-form fluxes so
that $\sum m_i = m\neq 0$. This is well understood for the mass-deformed
${\cal N}=4$ theory from the point of view of the AdS/CFT
correspondence \cite{polstr}. 

\subsection{Massive vacua}

For the class of deformations discussed in Section \ref{one}, the IIB
setup has only 3-form fluxes switched on.
As in the case of the mass deformed ${\cal N}=4$ theory, these 
fluxes would be expected to polarize the fractional D3-branes at the orbifold. 

In particular, with $\mu_i=\mu$
and $m_i=m$, the orbifold singularity remains unresolved, and we find that the
collection of $Nk$ fractional D3-branes blows up into $k$ fuzzy
spheres (or ellipsoids). This follows directly from the 
the weak-coupling or classical analysis of the Higgs vacuum
configurations where the (non-commuting) positions of the fractional
D3 branes satisfy the equations
\EQ{
{\Phi_r^2\over m^2}+{1\over {2\mu m}}
(\Phi^+_r\Phi^-_r+\Phi^-_r\Phi^+_r)={(\ell_r^2-1)\over 4};
\quad r=1,2\ldots k.
\label{fuzzy}
}

In the Higgs vacuum these fuzzy spheres must be thought of as $k$ concentric 
D5-branes with world-volumes ${\mathbb R}^4\times S^2$ with a
nontrivial action of the orbifold symmetry on the two-sphere. The dimension
$\ell_r$ of each block $\Phi_r,\Phi^\pm_r$ in the solution
Eq.(\ref{Higgsol}) gives the number of fractional branes constituting
the polarized D5-brane described by Eq.(\ref{fuzzy}).
It follows from the classification of Higgs vacua in Section 3.2
that each of the $k$ concentric spheres carries a
{\it distinct} fractional D3-brane charge. This is obvious in the
$k=2$ case which we discuss in more detail below. 
The net 3-brane charge carried by the $k$ concentric spheres is of
course $N$.  

In the large $N$ limit, if we look at massive vacua where the 
dimension $\ell_r$ of each block is chosen to scale as some positive
power of $N$, the D5-spheres become $k$ smooth spheres
embedded in
the six-dimensional space transverse to the D3-branes so that two
antipodal points intersect the $(x^4,x^5)$ fixed plane and are fixed
under the orbifold action. We remark that the Myers spheres we have
just described live in the flat orbifold, as our analysis of the Higgs
vacuum is classical, {\it i.e.\/}~$g^2_iN/4\pi\ll 1$. The fact that each
of the polarized spheres carries a distinct fractional brane charge
implies that they must each (non minimally) wrap topologically
distinct two-cycles at the orbifold singularity. 

The Type IIB configuration for vacua in other phases can be obtained
by the action of $\SL(2,\mathbb Z)$ duality on the above setup. For
example the confining vacua would be described by $k$ concentric
NS5-branes each carrying a distinct fractional D3-brane charge.

In the large $N$, $g_sN\gg 1$ limit, we expect the qualitative picture
of the concentric D5-brane spheres to survive as in the string dual of
the mass-deformed $\N=4$ theory of Polchinski and Strassler. The string
dual of the $A_{k-1}$  conformal quiver theory is the Type IIB theory
on $AdS_5\times S^5/{\mathbb Z_k}$ \cite{kachsilv} where the orbifold action
leaves
fixed a great circle $S^1\subset S^5$. The $\N=1^*$ deformation will
result in a warped geometry where the $k$ concentric 5-brane spheres will wrap
an $S^2\subset S^5$ intersecting the fixed circle at two antipodal
points. In a generic massive vacuum the 5-brane spheres will
have different radii and are thus expected to sit at different
values of the $AdS$ radial coordinate. Furthermore since each of the
5-branes carries a fractional 
D3-brane charge, normalizable modes for twisted sector fields
corresponding to VEVs in the gauge theory will be
turned on. 

\subsection{$\SU(N)\times\SU(N)$ $\N=1^*$ quiver theory}

We saw in Section 3.1 that the $\SU(N)^2$ theory has $N$ Higgs vacua where the
gauge
group is completely broken. In the basis that we have chosen, the
scalar field VEVs in these vacua 
decompose into 2 irreducible blocks, each forming a representation of
the $\SU(2)$ algebra. The dimensions $\ell_1$ and $\ell_2$ of these
blocks are odd integers satisfying $\ell_1+\ell_2=2N$. 

Thus, in each
Higgs vacuum of the ${\cal N}=1^*$, $A_1$ quiver theory, the $2N$
fractional D3-branes polarize into 2 concentric D5-brane spheres with radii 
given by the dimensions of the blocks. Importantly, since each
D5-sphere is 
constructed from an odd number ($\ell_i$) of fractional
D3-branes, it must necessarily carry a fractional D3-brane
charge. 
For example, in the case with $\ell_1=1$ and $\ell_2=2N-1$,
there is one fractional D3-brane stuck at the origin which can be
viewed as a D5-brane wrapping the collapsed $S^2$-cycle at the
orbifold. In addition there is a polarized D5-brane carrying $N-1$
units of whole D3-brane charge and one unit of fractional D3-brane
charge. In the large $N$ limit the smooth D5-sphere must wrap the
collapsed $S^2$ non minimally to account for the fractional D3-brane charge.

Confining vacua are obtained by the action of $S$-duality on the
above picture, turning it into 2 wrapped NS5-branes each carrying the
appropriate fractional D3-brane charge.

It is also worth noting that our classical solutions for   
the Higgs vacua with $\ell_1\neq\ell_2$, {\it i.e.}~vacua with unequal
sized D5-spheres have non-zero classical VEVs for twisted sector fields of the
form 
$\Tr\phi_1^2-\Tr\phi_2^2$ where $\phi_1,\phi_2$ are the adjoint
scalars in the two gauge group factors. Vacua which are described by two equal
sized
concentric 5-brane spheres ($\ell_1=\ell_2$)
do not have twisted sector VEVs at the classical level although in
general they may in the quantum theory when the gauge couplings for
the two factors are chosen to be different.

In the large-$N$ limit the fuzzy spheres become smooth and are
D5-branes wrapped on $S^2$ cycles supported by flux. From our analysis
of fractional branes at the unresolved orbifold it is not entirely
clear how the polarized spheres relate to the $S^2$-cycles associated
to the orbifold fixed point. The way to approach this question is to
investigate the embedding of the D5-spheres in the resolved orbifold
geometry with $\mu_1\neq \mu_2$. We will not address this issue in
this note.

\section{Non-perturbative effects in $\N=1^*$ theory}

To understand the origin of non-perturbative effects in $\N=1^*$
theories from a D-brane point of view we begin by exploring the
physics of the Higgs vacua. The resulting picture can then be extended
to the confining vacua by $S$-duality.  We expect that the physics of the Higgs
vacua can be completely understood in the limit of weak coupling where
all gauge couplings are taken to be small. In the semiclassical
limit, nontrivial physics in the holomorphic sector (in
particular, holomorphic in the gauge couplings) of the theory 
arises via instantons in each gauge group factor.
As usual instantons in the 4D gauge theory can be understood in the
D-brane language via D$(-1)$-branes, or D-instantons, in the presence
of D3-branes. 

We begin by illustrating how instanton effects in the Higgs
vacuum at weak-coupling can be understood in the simplest possible example,
namely the  
$\SU(N)$, $\N=1^*$ theory or the mass-deformed $\N=4$
theory. $M$-instanton contributions in this theory can be analyzed
via the $\U(M)$ gauge theory on the world-volume of $M$
D-instantons in the presence of the $N$ D3-branes. 

The world-volume
theory of D-instantons in the presence of $N$ coincident D3-branes
(in the absence of fluxes leading to the $\N=1^*$ deformation) is a
$\U(M)$ gauge theory with 8 supercharges \cite{inst}. It  
consists of 6 adjoint-valued real scalars which can be packaged into 3
complex scalars $\chi_i$ ($i=1,2,3$)
parameterizing  
the 6 coordinates transverse to both the D-instantons and 
the D3-branes in addition
to 4 adjoints $a_n$ corresponding to directions transverse to the
instantons but tangent to the 
world-volumes of the D3-branes. In the presence of the D3-branes
only 8 supercharges on the D-instanton gauge theory are preserved due
to fundamental strings stretched between the two types of branes
giving rise to $N$ fundamental hypermultiplets
$w_u ^a,\tilde w^u_b$, $(u,v=1,2,\ldots,N$ and
$a,b =1,2,\ldots M)$ transforming in the $(M,\bar M)$ of the $\U(M)$ gauge
symmetry and in the $(\bar N, N)$ of the $\U(N)$ flavor symmetry of the
D-instanton gauge theory. In the absence of $N=1^*$ deformations the
D-instanton gauge theory contains a term in the scalar potential:
\EQ{V_{\rm{D(-1)}}\thicksim \sum_{i}\bar{w}_a^u
({\bar\chi}_i)^a_b(\chi_i)^b_c w^c_u
+\sum_{i}\tilde w_a^u
(\chi_i)^a_b(\bar{\chi}_i)^b_c \bar{\tilde w}^c_u\ .}

We have not written the full scalar potential since only the terms
above and modifications to it will be relevant for the following discussion.
We also ignore terms in the D-instanton theory that are suppressed in
the 
$\alpha^\prime\rightarrow 0$ limit
(for instance, quartic couplings of the $\chi_i$'s which are
suppressed by two powers of $\alpha^\prime$ relative to the above terms).

Turning on the $\N=1^*$ deformation for the D3-brane theory has a
two-fold effect on the D-instanton gauge theory. Firstly, the $\N=1^*$
deformation introduces 3-form fluxes in spacetime which appear as masses for the 
adjoint scalars $\chi_i$. This effect is of the same order as the
quartic coupling of the $\chi_i$'s, {\it i.e.\/}~it is down by two
powers of $\alpha^\prime$. However, there is a
second effect that does contribute in the $\alpha^\prime\rightarrow 0$
limit. This is an effect {\it induced} by the
noncommutative positions of the D3-branes. The key point 
is that the D3-brane positions appear as hypermultiplet mass parameters
in the D-instanton gauge theory, by a modification of the couplings:
\EQ{\chi_i w \rightarrow \chi_i w+ w\VEV{\Phi_i} ;\;\;\;\;\;\
\tilde w \chi_i\rightarrow \tilde w \chi_i + \VEV{\Phi_i}\tilde w.
\label{dinst}}

 In the Higgs vacuum of the mass-deformed $\N=4$ theory the $\Phi_i$ in 
Eq. (\ref{dinst}) form an $N\times N$ irreducible representation of the
$\SU(2)$ algebra. These turn into mass matrices for the
hypermultiplets on the D-instanton theory leading to the modified F-term
equations:
\EQ{\left[\chi_i\right]_{M\times M}\left[w\right]_{M\times N}
=\left[w\right]_{M\times
N}\left[\Phi_i\right]_{N\times N}
\label{dinstb}}
and similar equations for $\tilde w$.

It is now fairly straightforward to see that these equations restrict
the values of $M$ for which $w,\tilde w$ are non-vanishing. 
When $M=1$, the $\chi_i$ are
numbers and Eq. (\ref{dinstb}) has no nontrivial solutions since
the matrices $\Phi_i$ are not simultaneously
diagonalizable. Similarly, it is easily seen that for any $M<N$, the
Eq. (\ref{dinstb}) has no nontrivial solutions unless the $\Phi_i$
are simultaneously block diagonalizable which is not possible since they
are irreducible $N\times N$ representations of the $\SU(2)$
algebra. In fact, nontrivial vacuum solutions exist only when $M$ is
an integer multiple of $N$, $M=Nn$ and $\chi_i$ are of the form:
\EQ{\left[\chi_i\right]_{Nn\times Nn}=
\MAT{\left[\Phi_i\right]_{N\times N}
&&&\\ &\left[\Phi_i\right]_{N\times N}&&\\%
&&\ddots&\\
&&&\left[\Phi_i\right]_{N\times N}}\label{dinstvev}}
Thus we find that in the Higgs vacuum of the mass deformed $\N=4$
theory D-instantons can sit on the $N$ D3-branes only in multiples of
$N$.
Thus the superpotential of the $\N=1^*$ theory in the
Higgs vacuum at weak coupling must naturally have the form
\EQ{W\thicksim \sum_n c_n \;{\rm exp}(2\pi iNn\tau).}

Ignoring vacuum-independent additive constants the
exact superpotential for the mass-deformed $\N=4$ theory has precisely
such an expansion at weak-coupling. This exact superpotential was
found in \cite{nick} and its value in the Higgs vacuum is given by
\EQ{W\thicksim \mu m^2\left[E_2(\tau) - NE_2(N\tau)\right],} 
where $E_2$ is the second Eisenstein series \cite{koblitz}. It turns
out that the first term proportional to $E_2(\tau)$ is simply a
vacuum-independent additive constant \cite{nickprem, oferandus} that
we ignore for the moment. Both in the weak-coupling or semiclassical
regime with Im$(\tau)\gg1$, and in the large-$N$ limit with 
${\rm Im} (N\tau) = N/g_s\gg1$ 
which is also the SUGRA regime, this superpotential
has an instanton  expansion   
\EQ{W\thicksim \mu m^2 \sum_n c_n {\rm exp}(2\pi iNn\tau)
+({\text{vacuum-independent terms}})}
with $c_n=\sum_{d|n}d$. This is precisely the expansion indicated by
our weak-coupling kinematical argument for contributions from 
D-instantons sitting on and tracking the polarized D3-branes. 

We now comment on the vacuum-independent additive constant $E_2(\tau)$
that has a natural weak-coupling expansion in terms of arbitrary
powers of instantons. These can be understood as arising from point-like
instantons which are separated from the D3-branes. These correspond to
solutions to Eq.~(\ref{dinstb}) that we overlooked. The point is that
if $w,\tilde w$ vanish there is no constraint on $M$ so that
arbitrary numbers of D-instantons can contribute. In this case
$\chi_i$ are not constrained and the D-instantons are not bound to the
D3-branes. Intuitively, they are not sensitive to the positions of the
D3-branes, the VEVs, and so their contribution is vacuum
independent. However, what is not so clear is why there couldn't be
contributions from mixed configurations of regular instantons bound to
the D3-branes and point-like instantons. We suspect that their absence
is due to extra unlifted fermion zero modes corresponding to the
superpartners of the 
relative separation of the two types of instantons. Integration over
the associated Grassmann collective coordinates would then yield
nothing. 
 
\subsection{World-sheet instantons from D-instantons}

A rather interesting outcome of the above picture is that we can now 
directly
attribute non-perturbative effects in the $\N=1^*$ theory to
world-sheet instantons (even at weak coupling). 
This follows from Eq.(\ref{dinstvev}) which demonstrates that in the
Higgs vacuum of the mass deformed $\N=4$ theory, instanton
contributions occur in multiples of $N$, where the 
associated D-instantons acquire non-commuting positions. In
particular, positions of the $N$ D-instantons in each of the $n$ groups track
the non-commuting positions of the polarized D3-branes. This implies
that each group of $N$ D-instantons has polarized into a higher
dimensional object, namely a D-string world-sheet wrapping the fuzzy
two-sphere. The fact that this object is a D-string follows from the
arguments in \cite{myers} according to which, D$p$-branes spreading out
into a transverse two-sphere via non-commuting positions 
naturally couple to $p+4$-form fluxes and must be thought of as
wrapped D$(p+2)$-branes.
The action associated to such a wrapped world-sheet is
given simply by the sum of the actions of the constituent
D-instantons, namely  ${\rm exp}(2\pi i N\tau)$. The solution  in
Eq.(\ref{dinstvev}) thus describes a D1-string world-sheet wrapped $n$
times around the expanded D5-brane with action $\exp(2\pi i n N\tau)$.

Although our arguments belong in a regime that is {\it a priori}
complementary to the SUGRA regime, the resulting picture coincides  
with that described in the large-$N$ string dual of
Polchinski and Strassler \cite{polstr}. It is a nontrivial fact that
the computation of the wrapped world-sheet action in the string dual as
the tension times the proper area of the Myers sphere gives the same
answer as the weak-coupling picture described above.

$\SL(2,{\mathbb Z})$ duality of the IIB theory (or the $\N=4$ theory)
permutes the phases and vacua of the $\N=1^*$ theory and the
superpotential terms in any vacuum can be attributed to $(p,q)$-string
instantons wrapping the two-spheres of polarized $(p,q)$ 5-branes. In
particular, in the
confining vacuum $S$-duality yields an expansion (for $g_sN \gg 1$) in powers of 
$\exp(-2\pi i {N\over\tau})$ $=\exp (-2\pi g_sN)$ which must be
associated with F-string instantons wrapping the two-cycle of a
polarized NS5-brane. This appears to be the right way to think
about non-perturbative effects in the confining vacuum in the regime of
large `t~Hooft coupling or the SUGRA regime as indicated in
\cite{nickprem,polstr}. 

\section{Non-perturbative effects in the $\N=1^*$ quiver theory}

We now extend the arguments developed in the previous section to the
$\N=1^*$ quiver models in a straightforward way and subsequently
compare with the associated, exact superpotentials.  

As before we will focus on the $\SU(N)\times\SU(N)$ theory. This theory
has $N$ Higgs vacua as explained earlier. Non-perturbative effects in
these vacua due to instantons in each gauge group factor can be
understood by thinking in terms of a $\U(M_1)\times\U(M_2)$ quiver gauge theory
that lives on the world-volume of $M=M_1+M_2$ fractional D(-1)-branes at
the orbifold in the presence of $2N$ fractional D3-branes. 
As usual the first group of $M_1$ fractional D-instantons
consists of D1-world-sheets wrapping the collapsed cycle at the
orbifold singularity and with action  $\exp(2\pi iM_1\tau_1)$. 
The second group may be thought of as wrapped 
anti-D1-world-sheets with action $\exp(2\pi i M_2\tau_2)=
\exp(2\pi i M_2(\tau-\tau_1))$ where $\tau$ is the overall gauge
coupling defined in Eq.(\ref{overall}) and maps onto the coupling of
the IIB string.

In the absence of the $\N=1^*$ deformations, the $\U(M_1)\times
\U(M_2)$ gauge theory on the D-instantons is an 8 supercharge theory
consisting of two complex 
adjoint scalars each living in one of the gauge group factors and 
parameterizing the positions of the D-instantons along the $(x^4,x^5)$
plane. In addition there are bi-fundamentals that parameterize their
positions in the orbifold directions. As in the 4D quiver gauge
theory, these can be amalgamated into three $(M\times M)$ matrices
(where $M=M_1+M_2$), $\chi$ and $\chi^\pm$ which contain the adjoints and
the bi-fundamentals respectively. As in Section 3, we will choose an
ordering of elements where odd-numbered rows and columns will be
associated with the first gauge group factor while even numbered rows
and columns will transform under the second 
factor of the D-instanton theory. Strings stretching between each
group of fractional D3-branes and the D-instantons associated with
that group give rise to $N$
flavors of fundamental hypermultiplets for each gauge group factor of  
the D-instanton gauge theory. These can be combined into two 
matrices $w^a_u,\tilde w_b^v$ with $u,v=1,2,\ldots
2N$ and $a,b=1,2,\ldots M=M_1+M_2$.

In the Higgs vacua the $2N$ fractional D3-branes polarize into two
concentric D5-branes each made up of an odd number $\ell_i$ of
fractional D3-branes with $\ell_1+\ell_2=2N$. 
The non-commuting VEVs of
the fractional D3-branes given by $\Phi, \Phi^\pm$ enter as mass parameters
for the fundamental hypermultiplets in the D-instanton theory. The
vacuum equations of the D-instanton gauge theory then require:
\EQ{\left[\chi\right]_{M\times M}\left[w\right]_{M\times 2N}
=\left[w\right]_{M\times
2N}\left[\Phi\right]_{2N\times 2N}\label{dinstc}}

We recall that in the Higgs vacua of the $\SU(N)^2$, $\N=1^*$
theory, the matrices $\Phi, \Phi^\pm$ consist of two irreducible
blocks of dimensions $\ell_1$ and $\ell_2$, each forming a
representation of the $\SU(2)$ algebra (and hence describing two
dielectric D5-branes each carrying a fractional D3-brane charge). 

We look for solutions where the D-instantons sit on the polarized D3-branes
({\it i.e.\/}~with non-vanishing VEVs for the fundamental
hypermultiplets). Following arguments identical to those in the preceding
section, we find that non-trivial solutions exist only when the
matrices $\chi,\chi^\pm$ are block diagonal and consist of $k_1$
copies of the $\ell_1$-dimensional representation of the $\SU(2)$
algebra and $k_2$ copies of the $\ell_2$-dimensional representation,
$k_1$ and $k_2$ being arbitrary integers.

The instanton contributions from such configurations can be enumerated
as follows. The polarized D5-brane with $\ell_1$ fractional branes
contains $(\ell_1+1)/2$ fractional D3-branes of the first type and
$(\ell_1-1)/2$ of the second type. Similarly the second Myers D5-brane
contains $(\ell_2-1)/2$ D3-branes of the first kind and $(\ell_2+1)/2$
of the second kind. Hence the net instanton contribution from the
above configurations is expected to be proportional to
\SP{&\exp\left(k_1 \left[2\pi i\tau_1 {\ell_1+1\over2}+2\pi i\tau_2
{\ell_1-1\over 2}\right]
\right)\times\exp\left(k_2
\left[2\pi i\tau_1{\ell_2-1\over2}+2\pi i\tau_2{\ell_2+1\over 2}\right]
\right)\\
&=\exp\left[2\pi i\tau_1(k_1-k_2)
+2\pi i\tau\left[Nk_2+(k_1-k_2){\ell_1-1\over 2}\right]
\right]\ .
\label{dstring}}
Here $\ell_1=1,3,5,\ldots 2N-1$ sweeps out the $N$ Higgs vacua of the
$\N=1^*$ quiver model. It must be
emphasized that these arguments are purely kinematical and it is by no
means necessary that all the contributions allowed by this mechanism
will actually appear in superpotential terms. Furthermore, as in the
case of the mass deformed $\N=4$ theory, point-like instantons ({\it
i.e.\/}~those that do not actually sit on the D3-branes and have
vanishing hypermultiplet VEVs) can and do contribute alongside the
above terms.

What appears to be significant in the context of the above mechanism,
is that non-perturbative effects at weak coupling in the Higgs vacua
can again be
directly thought of as world-sheet instanton effects. This
interpretation is due to the fact that solutions to Eq.({\ref{dinstc}})
describe $k_1$ wrappings of a D1-string world-sheet polarized from $k_1\ell_1$
fractional D(-1)-branes sitting on one polarized D5-brane; and $k_2$ wrappings
of a
D1-string world-sheet on the second polarized D5-brane. The
D-instantons track the non-commuting positions of the fractional
D3-branes on which they sit leading to an induced polarization of
D1-strings from fractional D-instantons. Not surprisingly, each of
these wrapped D-strings carries a distinct fractional D-instanton charge. 
In the confining vacua one expects
F-string world-sheet instantons to account for the non-perturbative
phenomena. We show below that analysis of the exact superpotential in
these vacua provides rather compelling evidence for this interpretation.

\section{The exact superpotential and wrapped world-sheets}

The exact superpotentials for the $A_{k-1}$, $\N=1^*$ quiver models
were derived in \cite{us}. For the sake of simplicity we restrict
attention to the $k=2$ case in the Higgs and confining vacua. These 
superpotentials turn out to be modular functions of the overall gauge
coupling $\tau$ 
(also the Type IIB coupling) reflecting $\SL(2,\mathbb Z)$ duality in
$\tau$ of the parent conformal quiver theory (accompanied by
appropriate re-scalings of the individual gauge couplings)
\cite{Lawrence:1998ja}. This $\SL(2,\mathbb Z)$ 
duality acts on the vacua of the mass-deformed theories by permutation
thus exchanging the phases of these theories. In particular $S$-duality
on the Higgs vacua yields confining vacua. This structure is encoded
in the condensates in each vacuum and in particular, in the value of the
superpotential in a given vacuum. These features are more or less
identical to those found in the context of the mass deformed $\N=4$
theory \cite{polstr,nick,nickprem,donwitt}.

In addition to $\SL(2,\mathbb Z)$, the $\N=2$ conformal quiver theory has
extra duality symmetries \cite{us, wittm, Lawrence:1998ja}. Part of
this duality symmetry is easy to 
understand from the gauge theory viewpoint as it simply corresponds to
shifting the theta angles of one of the gauge groups by multiples
of $2\pi$, $\tau_i\rightarrow \tau_i+n$. The additional symmetry is
visible from the associated string theory setups.
In the IIA/M-theory construction of
the elliptic models of \cite{wittm} this duality symmetry can be
understood as translations of the NS5-branes (as explained in
\cite{us}) by periods of the spacetime
torus which leaves the intersecting brane setup unchanged. Periodic
motion along one of the cycles (the M-dimension) corresponds to shifts
of individual theta angles, while motion along the other cycle of the
spacetime torus leads to shifts in the gauge couplings $1\over g^2_i$.

In the IIB picture of D3-branes at the orbifold, the gauge couplings
are related to the IIB $NS$ and $R-R$ two-form potentials as,
\EQ{{\tau_1}={1\over
4\pi^2\alpha^\prime}\left(\tau\int_{S^2}B_{NS}+ 
\int_{S^2}C_{RR}^{2}\right);\;\;\;\;\;\;
\tau_1+\tau_2=\tau={i\over g_s}+C^{0}_{RR},}
where the integrals are performed over the collapsed $S^2$ at the
orbifold. The integrals $\int C^2_{RR}/2\pi\alpha^\prime$ and $\int
B_{NS}/2\pi\alpha^\prime$ are both periodic variables with period $2\pi$
corresponding to the periodicity of the theta angle and of the
Yang-Mills gauge coupling $1\over g^2_i$ itself. Upon switching on the
$\N=1^*$ deformation these symmetries of the parent conformal theory act on the
vacua of the mass deformed theory by permutation. In particular, by
performing shifts of the gauge coupling ${4\pi\over g^2_1} \rightarrow
{{4\pi\over g^2_1}+{n\over g_s}}$ or $\tau_1\rightarrow \tau_1+n\tau$ we can
clock around the $N$ Higgs vacua of the $\N=1^*$ quiver
theory. This structure is precisely encoded in the exact
superpotentials which were obtained in \cite{us}, through the
appearance of terms that are elliptic in the individual gauge
couplings.

We now quote the results of \cite{us} for the superpotential and 
condensates evaluated 
in the Higgs and confining vacua with, for simplicity, equal
hypermultiplet masses $m_1=m_2=m$. For  a general $\N=1^*$ deformation
the superpotential is simply    
\EQ{ W_{\rm eff}= -{1\over
g^2_{YM}}(\mu_1\VEV{\Tr\Phi_1^2}+\mu_2\VEV{\Tr\Phi_2^2}).}
It is convenient to express this as a linear combination of the condensates 
\SP{H_e& =\VEV{\Tr\Phi_1^2}+\VEV{\Tr\Phi_2^2}\ ,\\
H_o&=\VEV{\Tr\Phi_1^2}-\VEV{\Tr\Phi_2^2}
}
which are even and odd, respectively, under the exchange of the two $\SU(N)$
factors.

These condensates can be expressed in terms of elliptic Weierstrass functions 
(see Appendix and \cite{ww} for details)    
\SP{\wp(z)&=-\zeta^\prime(z)\ ,\\
Q(z)&=\zeta(z)-{\zeta(\omega_1)z\over\omega_1}.} 
with half-periods 
\EQ{\omega_1=i\pi;\qquad\omega_2=i\pi\tau\ .}
Here we have introduced the function $Q(z)$ which has the property
that it is quasi-elliptic rather than elliptic. The linear piece
precisely gets rid of an identical term appearing in a semiclassical
Im$(\tau)\gg 1$
expansion of $\zeta(z)$. In terms of these functions the condensates
determining the superpotential in  the Higgs vacua $s=0,1,\ldots,N-1$
are (up to vacuum-independent additive constants)
\AL{ 
H_e(s)\Big|_{\rm Higgs}&=\left\{ {-N^2\over
2}m^2\Big[E_2(\tau)-N\;E_2(N\tau)\Big]+{1\over
2Nm^2}H_o(s)^2-2s^2Nm^2\right.\nonumber\\
&\left.+2N^2m^2\Big[\wp(2i\pi\tau_1|\omega_1,\omega_2)-N\wp(2i\pi
\tau_1+2s\omega_2|\omega_1,N\omega_2)\Big]\right\}\ ,\label{higgsa}\\
H_o(s)\Big|_{\rm Higgs}&=
2Nm^2\Big[NQ(2i\pi\tau_1+2s\omega_2|\omega_1,N\omega_2)
-Q(2i\pi\tau_1|\omega_1,\omega_2)+s\Big]\ .\label{higgsb}}
Note that the (quasi-)elliptic functions that encode the vacuum dependence
{\it i.e.\/}~the $s$-dependence are (quasi-)elliptic on a torus with
complex structure $N\tau$. It is also important to note that both the
condensates, as functions of $2\pi i \tau_1$, are actually elliptic
(and not quasi-elliptic) on this 
torus with complex structure parameter $N\tau$. This property can be
understood from the viewpoint of the IIA/M-theory setup corresponding
to the massive vacua, wherein the
M5-brane is actually wrapped on a torus which is an N-fold cover of
the spacetime torus with complex structure $\tau$. This leads to a low
energy duality, namely $\tilde S$-duality \cite{oferandus,us}.

The index $s$ that labels the $N$ Higgs vacua is directly related to
the dimensions $\ell_i$ of the two irreducible blocks characterizing
the classical solutions for the Higgs vacuum in
Eq.(\ref{Higgsol}). In particular $s=(\ell_1-1)/2$.
Thus $s$ is also related to the number of fractional D3-branes
$\ell_1,\ell_2$ making up each of the two D5-spheres in the IIB
picture of the Higgs vacuum. For instance, $s=0
$ is the Higgs vacuum
where precisely one fractional D3-brane remains stuck at the origin,
while the remaining $2N-1$ fractional D3-branes blow up into a
dielectric D5-brane.   

At weak coupling $g_{YM}^2/4\pi=g_s\ll 1$ and $g^2_i/4\pi\ll 1$, the
condensates $H_o$ and $H_e$ in (\ref{higgsa}) and (\ref{higgsb}), and
hence the $\N=1^*$ superpotential have an 
expansion in terms of instantons 
in each gauge group factor and whole instantons. This expansion is
also valid in the large $N$ limit in the SUGRA regime with
${\rm Im}(N\tau)=4\pi N/g^2_{YM}=N/g_s\gg 1$. Using standard formulae for the
expansion of Weierstrass $\wp$-functions and the Weierstrass
$\zeta$-functions \cite{ww}, and ignoring the vacuum independent
additive contributions we find that the
condensates $H_o$ and $H_e$ have a semiclassical expansion of the form,
\EQ{H_{o,e}(s)\Big|_{\rm Higgs}\thicksim\sum_{k_1,k_2}\;A_{k_1,k_2}
\exp(2\pi ik_1[s\tau+\tau_1])\exp(2\pi i k_2[(N-s)\tau-\tau_1]).}
The coefficients $A_{k_1,k_2}$ can be easily determined using standard
expansions and are completely vacuum-independent
constants.\footnote{These 
coefficients are simple integers, except for those
originating from the term $\propto H_o(s)^2$ in
Eq.(\ref{higgsa}). The latter term can be interpreted as arising from
an operator mixing with vacuum-independent coefficients that are
nontrivial functions of the gauge couplings (see \cite{oferandus,us} for
more comments on operator mixings).}
Note terms of this form are in complete agreement with the
contributions argued
in the previous section (see Eq.(\ref{dstring})) once we identify $s$
with $(\ell_1-1)/2$. Each of the terms in the above expansion has the 
interpretation of  
D1-string world-sheets multiply wrapping the two polarized D5-spheres, $k_1$ and
$k_2$ times
respectively. 

For every Higgs vacuum associated to a given $s$ there exists an
$S$-dual confining vacuum. Thus there are $N$ confining vacua that are
$S$-dual to the $N$ Higgs vacua. These confining vacua 
are related to each other by shifts of the theta angle of one of the
gauge couplings $\tau_1\rightarrow\tau_1+1$. Each of these confining vacua is
also a member of a
family of $N$ vacua related to each other by the shifts
$\tau\rightarrow \tau+1$. Hence there are $N^2$ confining vacua. 
The $S$-dual of the semiclassical instanton
expansion in the Higgs vacuum which is also a D-string world-sheet 
instanton expansion then naturally yields an F-string world-sheet
instanton expansion in the confining vacua. It is important to note
however that the latter expansion in terms of F-string instantons is
valid only in the regime of large `t Hooft coupling which is the SUGRA limit.

Without-loss-of-generality we now turn attention to the confining
vacuum $S$-dual to the $s=0$ Higgs vacuum.  The condensates in this
vacuum have the form,
\AL{ 
H_e\Big|_{\rm conf.}&=\left\{ {-N^2\over
2}m^2\Big[E_2(\tau)-{1\over N}\;E_2(\tau/N)\Big]+{1\over
2Nm^2}H_o^2\right.\nonumber\\
&\left.+2N^2m^2\Big[\wp(2i\pi\tau_1|\omega_1,\omega_2)-N\wp(2i\pi
\tau_1|N\omega_1,\omega_2)\Big]\right\}\ ,\label{conf}\\
H_o\Big|_{\rm conf.}&=2Nm^2\Big[NQ(2i\pi\tau_1|N\omega_1,\omega_2)
-Q(2i\pi\tau_1|\omega_1,\omega_2)\Big]}
Not surprisingly, the superpotential in the confining vacuum cannot be
interpreted in terms of smooth semiclassical configurations such as
instantons. This is because in the semiclassical limit the terms in
the superpotentials appear to have contributions with fractional
topological charge, often loosely termed as ``fractional
instantons'' (these are of course, distinct from fractional
D-instantons at the orbifold which correspond to objects with integer
topological charge residing in a given gauge group factor). In fact in
the limit Im$(\tau)\gg 1$ and Im$(\tau_1)\gg 1$, Eq.(\ref{conf}) has a
natural expansion in terms of powers fractional instantons in each
gauge group factor
\EQ{W_{\rm conf.}\thicksim\sum_{n,m=0}^{\infty} B_{n,m} e^{2\pi i n
\tau_1/N} e^{2\pi i m \tau_2/N}. 
} 
 
However, when we look at the confining vacuum in a SUGRA type limit,
{\it i.e.\/}~with $g_s N \gg 1$ a very different story emerges.  
The key point, which is very similar to that outlined in
\cite{nick, oferandus}, is that the superpotential in the confining
vacuum is actually a modular function of the variable $\tau/N$. This
is manifest in the term proportional to $E_2(\tau/N)$. However this is
also true for the terms involving the elliptic functions. The latter are
actually periodic on the torus with half-periods $N\omega_1=i\pi N$ and 
$\omega_2=i\pi\tau$ with complex structure parameter
$\tilde\tau=\tau/N$. Once again this 
is a consequence of the fact that holomorphic sector of the theory
possesses a low-energy duality symmetry which requires invariance under
modular transformations acting on $\tilde \tau$ -- namely, $\tilde{\rm
S}$-duality \cite{oferandus,us}.

In the
semiclassical limit Im$(\tau/N)=1/g_sN \gg 1$ the functions involved
have the natural expansion as above. However in the strong coupling
limit $g_sN\gg 1$ (setting $C^{0}=0$ for simplicity) it is natural to
perform the modular transformation $\tau/N \rightarrow
-N/\tau=ig_sN$ and re-expand in powers of $\exp(-2\pi iN/\tau)$. 
Under this transformation:
\SP{& E_2(\tau/N)\quad\longrightarrow \quad{N^2\over\tau^2}E_2(-N/\tau)
-{6N\over i\pi\tau}\ ,
\\& \wp(2\pi i\tau_1|N\omega_1,\omega_2)\quad\longrightarrow\quad
{N^2\over\tau^2}\wp({2\pi i N\tau_1\over
\tau}|N\omega_1,-N^2\omega_1/\tau)\ ,\\
& \zeta(2\pi i\tau_1|N\omega_1,\omega_2)\quad\longrightarrow\quad
{N\over\tau}\zeta({2\pi i N\tau_1\over
\tau}|N\omega_1,-N^2\omega_1/\tau)\ .
\label{wpconf}
}

Setting $C^0=0$, in the large $g_sN$ limit, $E_2(-N/\tau)=E_2(ig_sN)$ has an
expansion in powers of $\exp(-2\pi g_sN)$. Such a term can clearly be
obtained by $S$-duality on a configuration in the $s=(\ell_1-1)/2=0$ Higgs
vacuum
where
each D5-sphere is wrapped once by a D1-string world-sheet yielding an
action proportional to $\exp(-2\pi N/g_s)$ using
Eq.(\ref{dstring}). 

However the condensates in the confining vacuum encode an
additional important piece of information relevant to wrapped
world-sheets at the orbifold. Choosing all the theta angles to be zero
for simplicity we may rewrite Eq. (\ref{wpconf}) for instance as 
\EQ{\wp({2\pi i N\tau_1\over
\tau}|N\omega_1,-N^2\omega_1/\tau)=
\wp(iN {1\over 2\pi\alpha^\prime}\int_{S^2}B_{NS}|Ni\pi,-N^2\pi g_s).}
In the regime of large 't Hooft coupling $g_sN\gg 1$, this function can
be expanded in powers of $\exp(-2\pi g_s N)\times \exp(i n \int
B_{NS}/2\pi\alpha^\prime)$. The appearance of the phase factors in
this expansion is strongly suggestive of the fact that this is indeed
an expansion in powers of the F-string world-sheet instantons wrapping
the NS5-brane spheres  at the orbifold in the confining vacuum. Since the
5-branes themselves are
topologically (non-minimally) wrapped around the vanishing $S^2$ at the
orbifold, a string world-sheet wrapping such a 5-brane necessarily
picks up a phase contribution from the twisted sector field, namely
the integral of $B_{NS}$ over that two-cycle. In particular the
superpotential for the $s=0$ confining vacuum has the form
 \SP{W_{\rm conf.}\thicksim
\sum_{k_2,k_2} A_{k_1,k_2}\exp(-2N\pi k_2g_s +i{(k_1-k_2)\over
2\pi\alpha^\prime}\int B_{NS})}

Clearly these terms are consistent with a picture of $k_1$ F-string
world-sheets wrapping the first 5-brane sphere and $k_2$ wrappings of
the second 5-brane. The orientations of the wrappings are opposite
since each of the 5-branes carries a different fractional D3-brane charge.
This accounts for the fact that the there are phase
contributions proportional to $(k_1-k_2)\int B_{NS}$. This expansion
of the superpotential terms also implies that the polarized 5-branes
must be classified by the homology of the vanishing two-cycles
of the orbifold, {\it i.e.\/}~they must wrap (non-minimally) the $S^2$-cycles at
the
orbifold. We also remark that this world-sheet expansion of the
superpotential in the confining vacuum follows directly from an $S$-duality on
the expression in the Higgs vacuum. However it is important to note
that the expansion in terms of world-sheet instantons in the confining
vacuum is only valid in the regime of large `t Hooft coupling or the
SUGRA regime. At weak coupling it only appears to have an expansion in
terms of fractional instantons in each gauge group factor, which then
get re-summed into world-sheet 
instantons in the SUGRA limit.   

\section{Relevant deformation of the conifold theory}

One of the remarkable features of the analysis of the $\N=1^*$ quiver
theories is that the massive vacuum and phase structure is 
independent of the actual values of the mass parameters in the
deformation. In particular the condensates of chiral operators (in the
$\N=1$ sense) are independent of the specific choice of mass
parameters. One way to understand this is to view these theories as
softly broken $\N=2^*$ theories. From this viewpoint, the massive vacua
of the $\N=1^*$ theories are simply the maximally singular points
\cite{us} on the Coulomb branch of the unperturbed $\N=2^*$
theories and the condensates of chiral operators are given by the
locations of these  special singular points on the Coulomb branch
moduli space. The latter are of course independent of the choice of
$\N=1^*$ parameters. The low energy effective superpotentials for a
wide class of $\N=1^*$ theories can thus be automatically determined
as they  are simply special linear combinations of the chiral
condensates. 

Thus we may imagine a special $\N=1^*$ deformation of the
$\SU(N)\times\SU(N)$ quiver theory where we give masses
$\mu_1=-\mu_2=\mu$ to the two adjoint chiral multiplets: 
\EQ{\Delta W=-{1\over g^2_{YM}} \left(\mu\Tr\Phi_1^2-\mu\Tr\Phi_2^2+
m\Tr\Phi_{1,2}\Phi_{2,1}
+m\Tr\tilde{\Phi}_{2,1}\tilde\Phi_{1,2}\right).}
When $\mu\ll m$ this theory can be viewed as a soft
breaking of $\N=2$ SUSY Yang-Mills with massive bifundamental
hypermultiplets.  However, when $\mu\gg m$ we may also think of this
theory as a relevant deformation of the Klebanov-Witten conifold theory 
\cite{Klebanov:1998hh}. (With $m=0$, the bifundamental masses 
correspond to 
operators with dimension $3/2$ after accounting for the anomalous
dimensions at the infrared fixed point of the conifold theory.) The
low energy effective superpotential for this theory is simply
\EQ{W_{\rm eff}=-{\mu\over g^2_{YM}}(\VEV{\Tr\Phi_1^2}-\VEV{\Tr\Phi_2^2}).}
Recall that in the $m=0$ theory, one may integrate out the $\Phi_i$ to
obtain the quartic superpotential of the conifold theory. Thus the
effective superpotential must be odd under an exchange of the two
$\SU(N)$ factors. Using this and the appropriate semiclassical limit,
one finds from the results of \cite{us},\footnote{In \cite{us} 
the condensates were directly related to the
Hamiltonians 
of the spin generalisation of the Calogero-Moser integrable system. In this
context the 
Hamiltonian associated with the $H_o$ has precisely the right symmetry property
under the exchange and is hence identified with the expectation value of
$\VEV{\Tr\Phi_1^2}-\VEV{\Tr\Phi_2^2}$.}  
the following expression for
the exact superpotential of the theory:
\EQ{W_{\rm eff}\propto H_o.}
This is the exact low energy effective superpotential for the
deformation of the conifold SCFT by relevant operators corresponding
to mass terms for the bi-fundamentals of the orbifold theory. 

Note that this superpotential has a semiclassical instanton expansion
in the Higgs vacua which appears to satisfy the basic selection rules
described earlier.  However, the classical vacuum equations can no
longer be interpreted in terms of polarization of D3-branes into
5-branes. Instead, 5-branes automatically appear in the IIB picture
since the mass perturbation corresponds to a resolved orbifold with
blown up $S^2$ cycles and the fractional D3-branes are D5-branes
wrapped on this $S^2$. Non-perturbative effects in the gauge theory are
now simply world-sheet instantons wrapping this blown-up $S^2$.

\section{Conclusions}

In this note  we have discussed aspects of the Type IIB brane setup of
the $A_{k-1}$, $\N=1^*$ quiver models with gauge group $\SU(N)^k$ and
non-perturbative physics therein. These theories, obtained as relevant
deformations of the UV-finite $\N=2$ quiver theories, have vacua with
a mass gap where the theory is realized in Higgs and confining
phases. We argue that the IIB setup for these theories involves $k$
polarized 5-branes obtained from $Nk$ fractional D3-branes at the
unresolved orbifold in the presence of 3-form fluxes corresponding to
the mass deformation. The Higgs vacua are described by $k$ concentric
D5-branes with world-volume ${\mathbb R^4}\times S^2$, while the
confining vacua are described by $k$ concentric NS5-branes. In
addition each of these 5-brane spheres has a distinct fractional
brane charge resulting in the interpretation that they must each be
homologous to a distinct $S^2$-cycle of the resolved orbifold. The
non-perturbative effects in the Higgs vacua can be analysed at weak
coupling and are found to be due to D-instantons polarized to
Euclidean D-string world-sheets wrapping the polarized
D5-spheres. This is obtained by analysing the associated D-instanton
gauge theory. The latter yield selection rules for the from of
instanton contributions in the Higgs vacua.
The
physics of the confining vacuum is $S$-dual to this and originates from
wrapped F-strings. The exact superpotentials provide a nontrivial
confirmation of this picture. The strong coupling expansion of the
superpotential produces precisely the terms that would be obtained by
wrapping F-strings on the NS5-brane spheres at the orbifold. These
constitute the basic ingredients of the large-$N$ string dual of the
$\N=1^*$ quiver theories.

\vspace{1cm}

{\bf Acknowledgments:} We would like to thank N. Dorey for several
extensive discussions. We would also like to thank
B. Acharya, M. Douglas and R. Myers for useful comments.

\startappendix

\Appendix

In this Appendix we state some properties of the elliptic and modular
functions in the paper. For further details see \cite{us,ww}. The
Weierstrass $\wp$-function is an even function with periods
$2\omega_1\equiv 2i\pi$ and $2\omega_2\equiv2i\pi\tau$ defined via:
\EQ{\wp(z)={1\over z^2}+
\sum_{m,n\neq (0,0)}\left\{
{1\over{(z-2m\omega_1 -2n\omega_2)^2}} -{1\over{(2m\omega_1 +2n\omega_2)^2}}
\right\}.}
The Weierstrass $\zeta$-function is defined via
\EQ{\wp(z)=-\zeta^\prime(z).}
It is quasi-elliptic:
\EQ{\zeta(z+2\omega_{1,2})=\zeta(z)+2\zeta(\omega_{1,2}).}
with 
\EQ{\omega_2\zeta(\omega_1)-\omega_1\zeta(\omega_2)={i\pi\over 2}.}
The function $Q(z)=\zeta(z)-\zeta(\omega_1)z/\omega_1$ satisfies
\EQ{Q(z+2\omega_1)=Q(z);\qquad Q(z+2\omega_2)=Q(z)-1.}
Defining $\tau\equiv\omega_2/\omega_1$ and $q\equiv e^{2\pi i\tau}$, $\wp(z)$
has the following (instanton) expansion:

\EQ{\wp(z)=-{\zeta(\omega_1)\over\omega_1}-
{1\over 4}{\text {cosec}}^2({\pi z\over 2\omega_1})
+2\sum_{n=1}^\infty {nq^n\over{1-q^n}}\cos({n\pi z\over \omega_1}).}

\end{document}